\documentclass{SciPost}
\usepackage{etoolbox,xspace}
\newcommand{\ttz}{\ensuremath{\text{t}\bar{\text{t}}Z}\xspace}
\newcommand{\ttg}{\ensuremath{\text{t}\bar{\text{t}}\mathrm{\gamma}}\xspace}
\newcommand{\tgq}{\ensuremath{\text{t}\mathrm{\gamma}\text{q}}\xspace}
\newcommand{\twg}{\ensuremath{\text{t}\text{W}\mathrm{\gamma}}\xspace}
\newcommand{\ttbar}{\ensuremath{\text{t}\bar{\text{t}}}\xspace}
\newcommand{\ccbar}{\ensuremath{\text{c}\bar{\text{c}}}\xspace}
\newcommand{\bbbar}{\ensuremath{\text{b}\bar{\text{b}}}\xspace}
\newcommand{\pt}{\ensuremath{p_{\text{T}}}\xspace}
\newcommand{\stt}{\ensuremath{\sigma_{\ttbar}}\xspace}
\newcommand{\mtt}{\ensuremath{m_{\ttbar}}\xspace}
\hypersetup{
    colorlinks,
    linkcolor={red!50!black},
    citecolor={blue!50!black},
    urlcolor={blue!80!black}
}

\usepackage[bitstream-charter]{mathdesign}
\DeclareSymbolFont{usualmathcal}{OMS}{cmsy}{m}{n}
\DeclareSymbolFontAlphabet{\mathcal}{usualmathcal}

\fancypagestyle{SPstyle}{
\fancyhf{}
\lhead{\colorbox{scipostblue}{\bf \color{white} ~SciPost Physics Proceedings }}
\rhead{{\bf \color{scipostdeepblue} ~Submission }}

\fancyfoot[C]{\textbf{\thepage}}
}

\begin{document}

\pagestyle{SPstyle}

\begin{center}{\Large \textbf{\color{scipostdeepblue}{
TOP2024: an overview of experimental results 
}}}\end{center}

\begin{center}\textbf{
Abideh Jafari\textsuperscript{1$\star$,$\dagger$},
}\end{center}

\begin{center}
{\bf 1} Department of Physics, Isfahan University of Technology, Isfahan 84156-83111, Iran
\\[\baselineskip]
$\star$ \href{mailto:abideh.jafari@iut.ac.ir}{\small abideh.jafari@iut.ac.ir}\,,\quad
$\dagger$ \href{mailto:abideh.jafari@cern.ch}{\small abideh.jafari@cern.ch}
\end{center}

\definecolor{palegray}{gray}{0.95}
\begin{center}
\colorbox{palegray}{
  \begin{tabular}{rr}
  \begin{minipage}{0.36\textwidth}
    \includegraphics[width=60mm,height=1.5cm]{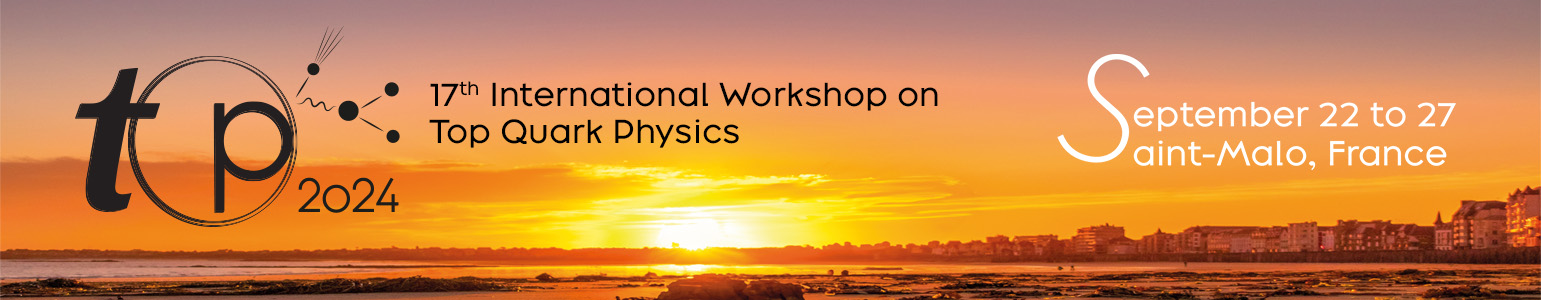}
  \end{minipage}
  &
  \begin{minipage}{0.55\textwidth}
    \begin{center} \hspace{5pt}
    {\it The 17th International Workshop on\\ Top Quark Physics (TOP2024)} \\
    {\it Saint-Malo, France, 22-27 September 2024
    }
    \doi{10.21468/SciPostPhysProc.?}\\
    \end{center}
  \end{minipage}
\end{tabular}
}
\end{center}

\section*{\color{scipostdeepblue}{Abstract}}
\textbf{\boldmath{%
The $17^\text{th}$ edition of the international workshop on top quark physics featured a diverse set of outstanding results. This note is an attempt to summarize the workshop from the experimental perspective and suggest ways forward for the future investigations. As it has not been possible to touch upon every single physics analysis in this note, interested reader are referred to individual proceedings for details.
}}

\vspace{\baselineskip}

\noindent\textcolor{white!90!black}{%
\fbox{\parbox{0.975\linewidth}{%
\textcolor{white!40!black}{\begin{tabular}{lr}%
  \begin{minipage}{0.6\textwidth}%
    {\small Copyright attribution to authors. \newline
    This work is a submission to SciPost Phys. Proc. \newline
    License information to appear upon publication. \newline
    Publication information to appear upon publication.}
  \end{minipage} & \begin{minipage}{0.4\textwidth}
    {\small Received Date \newline Accepted Date \newline Published Date}%
  \end{minipage}
\end{tabular}}
}}
}


\vspace{10pt}
\noindent\rule{\textwidth}{1pt}
\tableofcontents
\noindent\rule{\textwidth}{1pt}
\vspace{10pt}


\section{Introduction}
\label{sec:intro}
Since its discovery in 1995~\cite{CDF:1995wbb,PhysRevLett.74.2632}, the heaviest known particle, the top quark, has been a focus of studies in the experimental and theoretical particle physics. The large mass of the top quark results in a very short life time as well as a Yukawa coupling close to one. The particle is therefore an excellent candidate to study the bare quark properties, to understand the electroweak symmetry breaking, and to connect with new phenomena. In proton-proton (pp) collisions at the Large Hadron Collider (LHC), top quarks are abundantly produced in pair (\ttbar) through strong interactions, and to lesser extent, single via electroweak processes (single-top). The production in both cases can be accompanied by extra particles, including jets, vector bosons, the Higgs boson, and even additional top quarks. Currently, most of the production modes are observed and/or precisely measured. From precise measurements, the top quark mass and properties are determined while some other parameters of the standard model of particle physics (SM) such as the strong coupling constant, $\alpha_{\text{S}}$, can be extracted. At the same time, searches for new phenomena with top quarks together with the interpretation of top quark measurements have put strong constraints on theories beyond the standard model (BSM).

\begin{figure}[!ht]
    \centering
        
    \includegraphics[width=.55\textwidth]{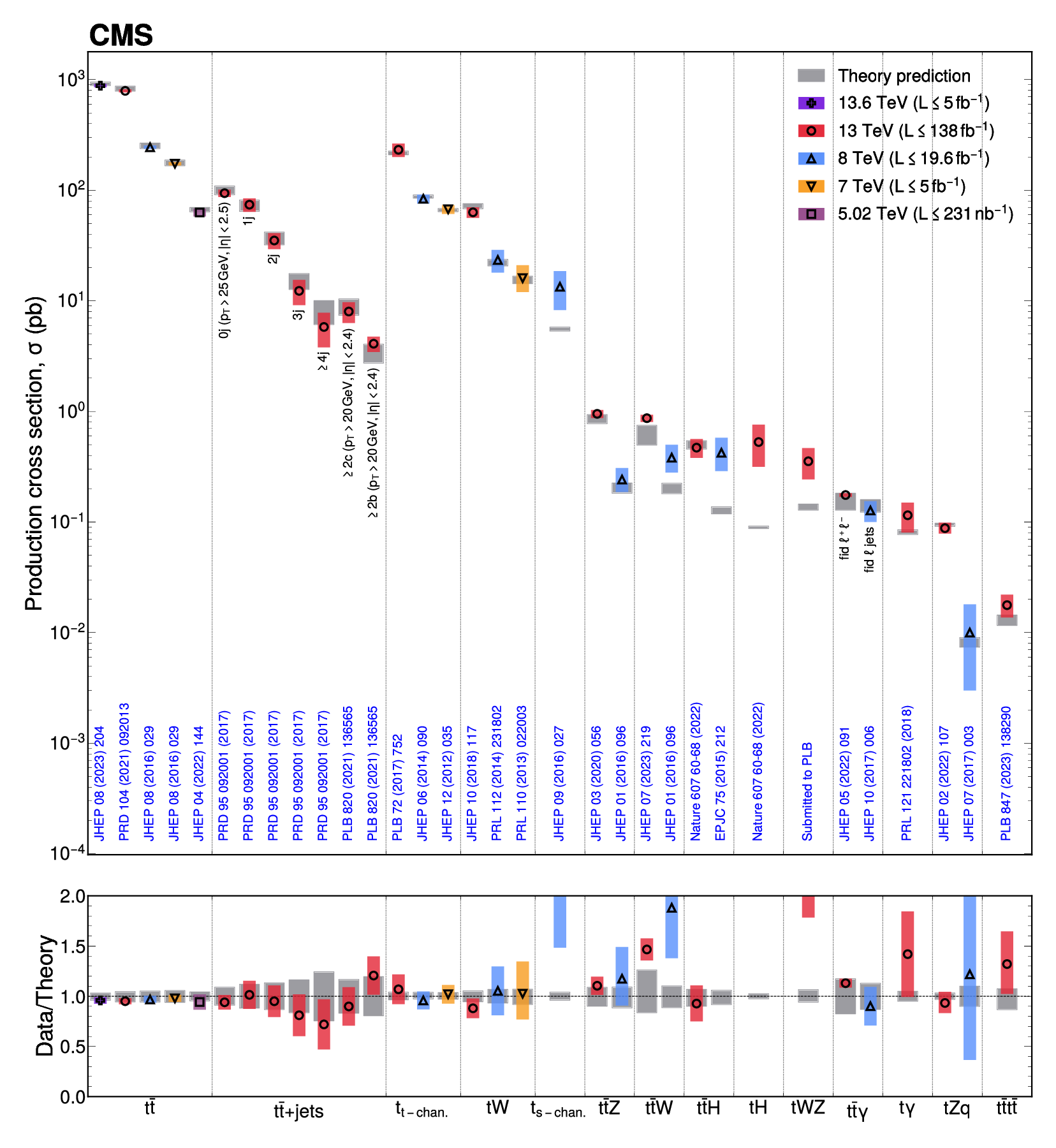}
    \caption{Summary of production cross section measurements involving top quarks by the CMS experiment~\cite{CMS:2024gzs}. Similar information is available in Ref.~\cite{ATL-PHYS-PUB-2024-006}, performed by the ATLAS collaboration.
    }
    \label{topsummary}
\end{figure}

The latest measurements and searches in top quark physics at the LHC rely on about two decades of scientific research, acquired knowledge, and innovation. The state-of-the-art analysis techniques and particle identification, theory advancements, and the large data sample provided by the LHC, have enabled the ATLAS and CMS experiments to perform high precision measurements, observe extremely rare processes, and strongly rule out various BSM possibilities. Figure~\ref{topsummary} shows the inclusive cross section measurements using pp collisions collected by the CMS experiment~\footnote{Similar information is available in Ref.~\cite{ATL-PHYS-PUB-2024-006}, performed by the ATLAS collaboration.}, covering few orders of magnitude for processes involving top quark(s)~\cite{CMS:2024gzs}. The majority of measurements are limited by systematic uncertainties. It must also be noted that associated to almost every results in Fig.~\ref{topsummary}, there are (multi)differential measurements, extraction of SM parameters, and interpretations in the context of BSM theories. 

Considering different top quark production modes, the large \ttbar cross section, \stt, has made it a unique target for the measurement of top quark properties, extraction of relevant SM parameters, and interpretations in the context of new theories. It was the first to be measured at the start of the LHC Run~3~\cite{ATLAS:2023slx,CMS:2023qyl}, already with a high precision. In addition, CMS reported a measurement at $\sqrt{s} = 5.02\,\text{TeV}$~\cite{CMS:2024ghc}, to compare with the corresponding ATLAS result~\cite{ATLAS:2022jbj} and to be used as a reference for top quark studies in heavy ion collisions at the same center-of-mass energy, $E_{\text{c.o.m}}$. The ATLAS collaboration presented a \stt measurements in proton-lead (pPb) collisions at $\sqrt{s} = 8.16\,\text{TeV}$~\cite{ATLAS:2024qdu} to compare with an earlier result by CMS~\cite{CMS:2017hnw}. New results from both experiments call for further discussions and comparisons of analysis details and simulation settings, which can be particularly interesting when heavy ions are involved.

The available knowledge and expertise paves the way for future investigations in the field of top quark physics. By the end of 2025, the full LHC data at $\sqrt{s}=13.6\,\text{TeV}$ will be available for physics analyses. The High-Luminosity LHC is expected to start in mid-2030, providing the largest-ever amount of proton-proton collision data. At the same time, experiments will undergo a substantial upgrade to sustain the immense luminosity and to fully exploit the wealth of the data. This is a unique opportunity to explore uncharted territories in top quark physics, drawing on the existing and developing knowledge. 

In addition to a summary of the experimental results presented in TOP2024, this note provides a few suggestions to expand the research in the coming years. As a disclaimer, the article presents a personal view and does not cover \textit{all} possibilities. 
\section{Machine learning for top quark measurements}\label{ml4top}
Modern machine learning (ML) techniques are nowadays an integrated part of experimental high energy physics. In top quark physics in particular, they play a central role at different levels of the physics analysis, from more precise identification of leptons and b-quark jets to advanced parts of the analysis design and its interpretations\footnote{See dedicated presentations in~[\href{https://indico.cern.ch/event/1368706/contributions/6137485/}{1}, \href{https://indico.cern.ch/event/1368706/contributions/6137486/}{2}, \href{https://indico.cern.ch/event/1368706/contributions/6137487/}{3}].}. Using sophisticated graph neural networks, the latest ATLAS jet flavor tagger has shown more than two\,(three) times rejection rate for c-quark\,(light-quark and gluons) jets while preserving 70\% efficiency for jets originating from b quarks~\cite{ATL-PHYS-PUB-2022-027}. The CMS {\sc UParT} algorithm has reached a similar performance~\cite{CMS-DP-2024-066} where for the first time, a strange-tagging node is also introduced in the network. Future improvements on the s-tagger can help identify the rare top quark decay mode $\text{t}\to\text{W\,s}$ and directly probe the $V_{\text{ts}}$ element of the CKM matrix.

Another example of the ML usage is to preserve the precision of top quark measurements where the limited size of Monte-Carlo (MC) samples, especially those with a varied parameter to assess the associated systematic uncertainty, is a limiting factor. Using ML techniques, the effect of the varied parameter can be estimated in the phase space of study with no need for large MC production~\cite{CMS:2024jdl}. This approach is crucial, especially for the high-luminosity LHC, where MC production can hardly cope with the amount of collected data for precise measurements. Other areas where ML has an impact on top quark physics include the reconstruction of the top quark event hypothesis, particularly in the presence of neutrinos and extra particles~\cite{CMS-PAS-TOP-24-001}, and estimation of backgrounds~\cite{CMS:2023zdh}. 
\section{Opportunities with \ttbar differential measurements}
\label{ttdif}
\begin{figure}[!ht]
    \centering
        
    \includegraphics[width=.85\textwidth]{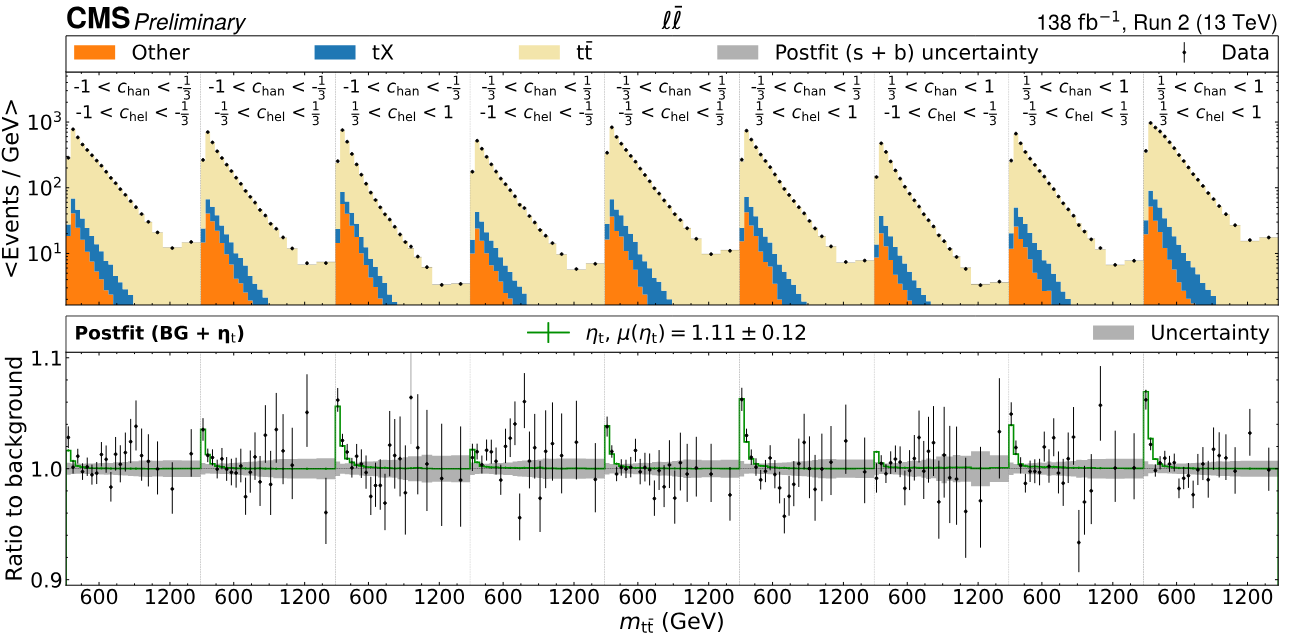}
    \caption{Observed and expected \mtt distribution in bins of spin correlation observables shown for the dilepton (e and/or $\mu$) final state summed over lepton flavors and analysis eras. In the first panel, the data points with statistical error bars and predicted perturbative QCD backgrounds (colored histograms) are compared before the fit to the data, and the corresponding prefit uncertainty is shown with a gray band. In the second panel, the ratio is shown with the best fit normalization applied considering toponium contribution in the fit. There, the gray band shows the postfit uncertainty, and the respective signal is overlaid with its best fit model parameters. Adapted from Ref.~\cite{CMS-PAS-HIG-22-013}.
    }
    \label{att}
\end{figure}

Differential distributions of \stt have been measured since the early times of the LHC data taking and provided valuable inputs to the upcoming measurements and to the theory community. The well-known observed slope in the differential measurement of top quark transverse momentum \pt with respect to theory predictions at the time triggered a chain of studies over different data-taking eras. Multidifferential measurements were performed to possibly localize the effect where at the same time, higher order calculations could explain (part of) the disagreements. The acquired knowledge from multidifferential measurements went beyond understanding the top quark \pt distributions and are used to extract SM parameters such as $\alpha_{\text{S}}$ and top quark mass, and to constrain the gluon density function of protons~\footnote{Refer to \href{https://indico.cern.ch/event/1368706/contributions/6011820/}{this} presentation for a comprehensive overview.}. The state-of-the-art ML methods have made it possible to perform simultaneous differential measurements of many more observables with a higher granularity in other physics processes~\cite{ATLAS:2024xxl,CMS-PAS-SMP-23-008}. It is desired to use the same techniques for differential \stt measurements. 
\paragraph{Possibility for a \ttbar bound state (\textit{aka} toponium)} Heavy pseudoscalars decaying to \ttbar can create a peak-dip structure in the \mtt distributions, if exist. In a search for such particles, using \ttbar angular and spin correlation observables, CMS has observed an excess with a significance of more than $5\sigma$ at \mtt values close to the \ttbar production threshold~\cite{CMS-PAS-HIG-22-013}. The excess is compatible with the presence of a toponium signal, using the available calculations in which the color-octet state is not accounted for. Figure~\ref{att} shows the observed and expected \mtt distributions in bins of spin correlation observable. The SM distributions with an injected toponium signal are fitted to and compared with the data in the lower panel. The pseudoscalar hypothesis is slightly favored by the data for the observed excess. A similar measurement by ATLAS together with more realistic theory predictions and proposals for complementary measurements seem necessary to better identify the nature of this signal. 
\paragraph{Quantum entanglement outside the light cone} Using the $\Delta\phi$ between the two leptons from W bosons in \ttbar decays, ATLAS and CMS experiments observed quantum entanglement between the two top quarks near the \ttbar production threshold~\cite{ATLAS:2023fsd,CMS:2024pts}. The entanglement signal remained statistically significant in the presence of possible \ttbar bound state. In another work~\cite{CMS:2024zkc}, the CMS experiment exploited the full \ttbar spin density matrix and observed entangled top quarks in part of the phase space with high \mtt and low top quark scattering angles. This is particularly important because in this regions top quarks are expected to be largely outside each other's light cone and the information cannot be transmitted with hidden variables~\footnote{Possibilities to experimentally probe quantum entanglement in single-top were presented \href{https://indico.cern.ch/event/1368706/contributions/6012495/attachments/2932848/5150748/LS.pdf}{here}.}. 
\section{Associated top quark productions}\label{ttxtx}
\begin{figure}[!ht]
    \centering
        
    \includegraphics[width=.75\textwidth]{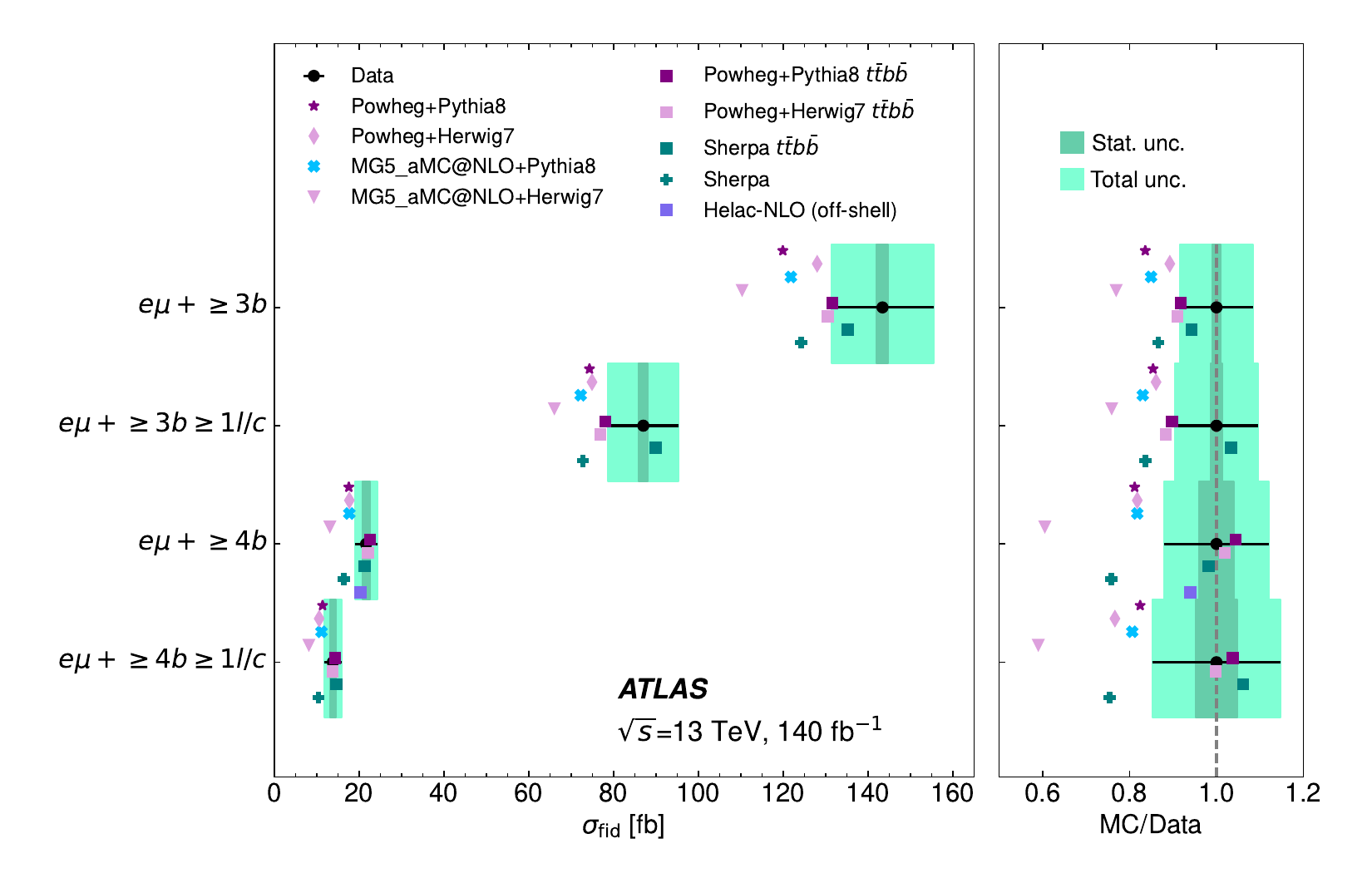}
    \caption{Measured fiducial cross sections compared with the central values of NLO {\ttbar}{\bbbar} predictions from different event generators and settings~\cite{ATLAS:2024aht}. The right hand panel shows the ratios of the MC predictions to the data. A similar trend is observed in the CMS measurement~\cite{CMS:2023xjh} performed in lepton+jets final states and compared with different event generators and settings.
    }
    \label{ttbb}
\end{figure}

While the \textbf{latest {\ttbar}{\bbbar} measurements} by ATLAS and CMS~\cite{ATLAS:2024aht,CMS:2023xjh} show similar trends in comparison with theory predictions, comparing of the results is not straight forward because of different final states and definitions of the fiducial phase space (see Fig.~\ref{ttbb} for the ATLAS result). In addition to harmonized selections, common MC samples with similar generators and scale choices would be extremely useful in this context. The CMS \textbf{{\ttbar}{\ccbar} measurement} has to be updated with the high-performance c-taggers, and in view of the latest ATLAS result~\cite{ATLAS:2024plw}. The \textbf{LHC searches for {\ttbar}{\ttbar}} have been concluded with the observation of the process in Run~2, with the possibility to combine the results of the two experiments with higher precisions in Run~3. The off-shell {\ttbar}{\ttbar} production has shown constraining power on the Higgs boson width and top-Higgs Yukawa coupling $\kappa_{\text{t}}$~\cite{ATLAS:2024mhs}. Given the sensitivity of differential \mtt and $\Delta y_{\ttbar}$ measurements to $\kappa_{\text{t}}$~\cite{CMS:2020djy}, there is more room for top quark physics to contribute to the Higgs sector of the SM.
The vector boson associated production of top quark involves {\ttbar}V, tVq, and tWV, with V here being a Z boson or a photon. Precise \ttg measurements exist where experiments need to harmonize the definition of the visible phase space, and the origin of the photon. The current \tgq observation has to be followed by differential measurements, preferably simultaneously with \ttg. To date, there are no dedicated \twg measurements at the LHC while an evidence has been found for the tWZ production~\cite{CMS:2023krq}. In view of future tWV measurements, the treatment of quantum interference with {\ttbar}V has to advance in simulation, ideally accounting for non-, single- and double-resonant processes at NLO QCD in a single MC sample. A simultaneous differential measurement of \ttz and tZq has been recently reported by CMS~\cite{CMS:2024mke}. It provides a consistent treatment of the two processes, which is particularly important for constraining top quark electroweak interactions in the context of the effective field theory (EFT). 

\section{BSM searches and interpretations}\label{bsm}
The top quark is a key player of BSM searches including dark matter, vector-like quarks, and lepton flavor and baryon number violations~\footnote{Details were presented in [\href{https://indico.cern.ch/event/1368706/contributions/6012427/}{1}, \href{https://indico.cern.ch/event/1368706/contributions/6012412/}{2}, \href{https://indico.cern.ch/event/1368706/contributions/6012521/}{3}]. }. In addition, there are huge efforts in both theory and experimental communities to constrain, with maximum information and minimum assumptions, the new couplings in the EFT Lagrangian using top quark measurements. Recently, CMS has released a global EFT fit based on likelihood functions from a wide range of measurements~\cite{CMS-PAS-SMP-24-003}. This can be confronted with the biggest global fit from the theory side~\cite{Celada:2024mcf} that constrains 50 Wilson coefficients using 455 data points. The experimental fit is going to expand with more event signatures, include latest modeling of the processes, and employ ML techniques. An additional step can be the inclusion of searches that are interpreted in the EFT framework~\cite{ATLAS:2024hac}. The theory fit can also improve by using more of the available differential measurements, especially those of associated top quark productions.

\section{Conclusion}
Top quark physics at the LHC advances relying on the outstanding work from theory and experiments. Exchanges across the communities are vital for the continuation of the research. On the experimental side, the succesful experience of common Monte-Carlo \ttbar samples between ATLAS and CMS can be extended to other top quark processes to facilitate the comparison of definitions and the results.

\section*{Acknowledgements}
The author thanks the organizers of the workshop for the invitation and for the excellent coordination. 

\paragraph{Funding information}
The participation of the author is financially supported by the Ministry for Research, Science, and Technology of Iran, and by Isfahan University of Technology.







\bibliography{SciPost_Proceedings_TOP2024_Template.bib}


\end{document}